\documentclass[prb,showpacs,twocolumn,aps,superscriptaddress]{revtex4}

\usepackage{graphicx}
\usepackage{bm}
\usepackage{amssymb}
\usepackage{amsmath}
\usepackage{subfigure}
\usepackage{tikz}

\begin{document}

\title{Doping on the kagome lattice: A variational Monte-Carlo study of the $t-J$ model}
\author{Siegfried Guertler}
\affiliation{Bethe Center for theoretical physics, The University of Bonn, Bonn, Germany}
\author{Hartmut Monien}
\affiliation{Bethe Center for theoretical physics, The University of Bonn, Bonn, Germany}
\date{\today }

\begin{abstract}
We study doping on the Kagome lattice by exploring the $t-J$-model with variational Monte-Carlo. We use a number of Gutzwiller projected spin-liquid and valence bond-crystal states and compare their energies at several system-sizes. We find
that introducing mobile holes drives the system away from the Spin-Liquid state proposed by Ran et al for the undoped system, towards 
a uniform state with zero-flux. On top of the uniform-state a VBC of the
Hastings-type is formed for low doping. The results are compared to exact diagonalization on small clusters. This agrees well.

\end{abstract}

\maketitle

\section{Introduction}

\begin{figure}
     \centering
\includegraphics[width=\columnwidth]{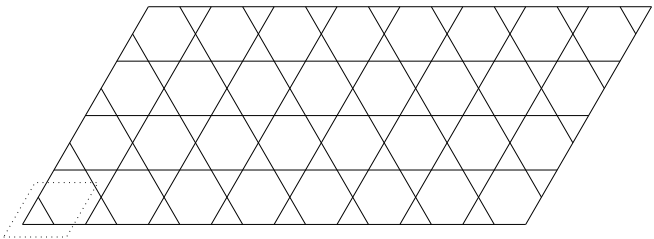}
\caption{The Kagome lattice consisting of corner sharing triangles. The 3-site physical unit-cell is indicated with dotted lines. The underlying Bravais lattice is triangular, leading to a hexagonal shaped Brillouin-zone as in the triangular lattice.}
\label{Kagome}
\end{figure}

Spin-liquids were first introduced in 1973 by P. W. Anderson as an alternative ground state to the anti-ferromagnet \cite{AN1}. These states were suggested for the high-$T_c$ superconductors (such as RVB-states) \cite{AN2}. Spin-liquids gained tremendous attention due to the discovery of a number of new materials such as ${\rm Na_4Ir_3O_8}$ \cite{okamoto1}, ${\rm SrCr_{8-x}Ga_{4+x}O_{19} }$ \cite{Ramirez2000,brohol1}, ${\rm ZnCu_3V_2(OH)_6Cl_2}$ \cite{shores1,ofer3,Helton1,Mendels1,Hiroi1} and ${\rm KFe_3(OH)_6(SO_4)_2}$ \cite{nishiyama1,Matan1,Imai1}, where the realization of such states is argued. These compounds share a Kagome like lattice in two or three dimensions, with the feature of corner sharing triangles (see Fig. \ref{Kagome}), being a prototype lattice for frustration with respect for the antiferromagnet. The Kagome lattice Heisenberg model has been studied extensively and the proposed ground states are a valence bond crystal \cite{HUS1}, 
a gapped spin-liquid state \cite{white} and a particular gapless spin-liquid state (dubbed U(1) Dirac state) \cite{RAN1,IQBAL}. 
Here the recent study by Yan, Huse and White found a gaped $Z_2$ spin-liquid state by DMRG, extrapolating remarkable well to the results from exact diagonalization,\cite{white} however a very recent rapid communication by Iqbal et al followed such states by VMC and found that the U(1) Dirac state, originally found by Ran et al still gives the best variational energy at half filling. While one may argue the validity of VMC due to the bias given by fixing a functional form of the wave-function, this can not be argued here, as the choice of the wave-function was in fact motivated by the DMRG study. One possible reason for this puzzling development, might be the difficulty of DMRG in two dimensions \cite{CHUNG1,CHUNG2}. We would like to emphasise here, that within the VMC the choice of the boundary condition changes the variational energy, but not the best variational state at half-filling.\\
Static impurities on this lattice have been studied by Dommange et al \cite{DOM}. This
study found the introduction of spinless holes leads to 
dimer-freezing. The interaction of the holes was found to be repulsive. A
single static or mobile hole injected in the Kagome and other frustrated
lattices has been investigated thereafter with exact diagonalization by
L\"auchli et al \cite{POIL} finding spin-charge separation and
dimer-freezing for the positive hopping sign of $t$. Particular the fillings
$n=1/3$ and $n=2/3$ in the $t-J$ model have been studied with mean-field calculations, exact
diagonalization and DMRG for a number of Kagome like lattices by Indegand 
  et al \cite{INDE}, proposing a ``bond-order-wave''-state as a generalization
of a ``valence bond solid". In a recent letter by Singh the Kagome lattice Heisenberg model with random vacancies is studied, and a ``valence-bond-glass'' state is argued \cite{SIN2}. The Hubbard model on the Kagome has been investigated with dynamic quantum Monte Carlo by Bulut et al \cite{BULUT} and strong short-range anti-ferromagnetic correlation was found by this study.\\
In this paper we study mobile holes in the Kagome lattice with the $t-J$-model and variational Monte Carlo for the first time. The question on doping is of very general interest in condensed matter physics. In addition very recently doping has been addressed experimentally on the pyrochlore lattice. This lattice consists of alternating Kagome and Triangular planes coupled in the direction perpendicular to the planes \cite{WANG}. In addition the highly degenerate nature of the Kagome lattice, makes it interesting to study different perturbation to the ground state. This has been pointed out recently by one of us in a preprint on the classical model \cite{Martin} and is a viewpoint shared by some of our colleagues \cite{JEONG}. Therefore our study has some relevance to the half-filled case too. For small thermodynamic doping we test spin-liquid and valence-bond crystal trial states previously proposed for the Heisenberg model on the Kagome lattice. We find that in the presence of a few holes the Dirac spin-liquid state is unstable and a zero flux (uniform spin liquid) state is formed, accompanied by a ``Valence Bond Crystal" (VBC). As will be seen below we find remarkable well agreement with exact diagonalization on small clusters, even we used the most naive boundary conditions (periodic boundary conditions).\\

\section{Model and Method}

 We study the $t-J$ model within the subspace of single occupied lattice sites on the Kagome lattice (see Fig. \ref{Kagome}):

\begin{eqnarray}
H&=&-t\sum_{\langle ij\rangle}\sum_{\sigma=\uparrow\downarrow}\mathcal{P_G}\left(c^{\dagger}_{i\sigma}c_{j\sigma}+ {\rm h.c.}\right)\mathcal{P_G}+ \nonumber \\
 & & +J\sum_{\langle ij\rangle}\left(\mathbf{S}_i\cdot\mathbf{S}_j-\frac{1}{4}n_i n_j\right)
\label{ham}
\end{eqnarray}

Here $c_{j\sigma}$ is the electron annihilation operator of an
electron with spin $\sigma$ on site $i$, $\vec S_i$ is the spin-1/2 operator at site $i$, and
$n_{i\sigma}=c_{i\sigma}^{\dagger} c_{i\sigma}$. The sum $\langle i,j \rangle$
is over the nearest neighbors (n.n.) pairs on the Kagome lattice. $\mathcal{P_G}=\prod_{i}(1-n_{i\uparrow}n_{i\downarrow})$ is the Gutziller projection operator enforcing the single occupied site constraint. We fix $J=0.4t$ in what follows. As wavefunction we consider a combination of previously suggested Gutzwiller projected MF-states which are classified by the fluxes through the triangles and hexagonal respectively, and two types of explicit VBCs. Such states are constructed from the mean-field hamiltonian:

\begin{equation}
H_{{\rm MF}} =
\sum_{i,j,\sigma=\uparrow\downarrow}\chi_{ij}c_{i,\sigma}^{\dagger}c_{j,\sigma}+{\rm h.c.}
\end{equation}

\begin{figure}
     \centering
\includegraphics[width=0.98\columnwidth]{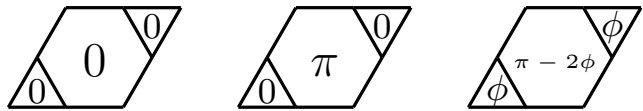}
\caption{The flux states U, D and C, being defined by the value of a flux when surounding a plaquette, as explained in the text.}
\label{UDC}
\end{figure}

\begin{figure}
     \centering
\includegraphics[width=\columnwidth]{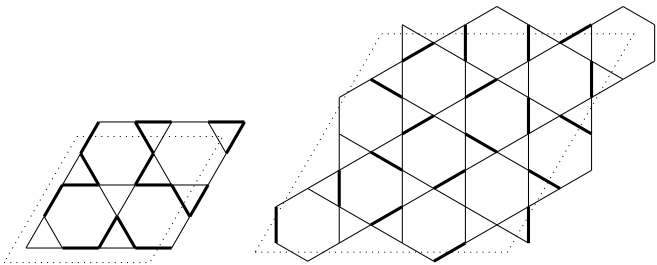}
\caption{Left: The patten of the VBC-1 state. The dotted lines indicate the 12-site unit cell. Right: The patten of the VBC-2 state. The dotted lines indicate the 36-site unit cell. The variational parameter $\chi$ tunes the ratio between the weak and strong bonds.}
\label{VBC12}
\end{figure}

This form can be obtained by rewriting the spin-operator in fermion operators and introducing the mean-field variable $\chi^*_{ij}=\langle c_{i,\sigma}^{\dagger}c_{j,\sigma} \rangle$.
One can define a flux $\phi$ through a given plaquette with $\exp{(\rm i\phi)}=\prod_{{\rm plaquette}}\chi_{ij}$ with $\chi_{ij}$ being defined on one edge of the plaquette. In addition to the phase $\phi$ on a plaquette, the amplitude of $\chi_{ij}$ can be defined on any bond, the latter defines a VBC. Choosing a mean-field unit-cell capable to accommodate the background-flux and VBC, one can write out this Hamiltonian for in matrix form. This matrix can then be diagonalized numerically to obtain the single-particle eigenfunctions. This being the basis for our many-particle wave-function in the VMC calculation, formulated in the usual form as the product of two Slater determinants (one for the up- and one for the down-spin electrons), where each element is one such single-particle eigenfunction. The Gutzwiller projection operator defined above is then applied on this state leading to the final VMC state to be optimized: $ | \Psi_{\rm VMC}(\chi_{ij})\rangle = \mathcal{P_G} | \Psi_{\rm MF}(\chi_{ij})\rangle $.
Projecting  is a crucial point, as this changes the mean-field energies dramatically. While this projector can not be treated properly in an analytic mean-field treatment, in VMC projection is exact.
The two VBCs we are choosing are the one suggested by Hastings \cite{HAS1} and the one suggested by Zeng and Marston \cite{Marston1}.  A particular flux state is parametrized with bond-modulation in the pattern of these two VBCs (see below), combining background-flux and VBC in one state. We use unit cells of the size of 12 and 36 sites, corresponding to smaller physical unit-cells of 3 and 9 sites respectively to accommodate these states. In the remainder of the paper we choose the following notation for the trial-states: 
\begin{enumerate} 
\item U: the uniform state with no flux through the triangles and hexagons, with a spinon Fermi-surface, \item D: the Dirac U(1) state, with zero flux through triangles and $\pi$-flux through the hexagonals, having two Dirac-points instead of a Fermi-surface, \item VBC-1($\chi$): the VBC state suggested by Hastings, with a 12-site unit-cell. the value $\chi$ in bracket parametrizes the strength of the bond-modulation, as the amplitude ratio between strong and weak bonds. \item VBC-2($\chi$): the VBC patten suggested by Zeng and Marston, requiring a 36 site unit cell. The value $\chi$ has the same meaning as above. \item C($\phi$): the Chiral state defined by the perturbation towards the D-state. Here $\phi$ labels the flux through the triangles, while the flux through the hexagons is $\pi-2\phi$.
\end{enumerate}
E.g. U+VBC-1(1.1) means therefore a state with uniform flux and a bond-order patten resembling the one of VBC-1 with a ratio of the strong compared to the weak bonds of $\chi=1.1$. Similarly we will label other states. The states U, D and C are visualized in Fig. \ref{UDC} the two VBCs can be found in Fig. \ref{VBC12}. All bonds belonging to the patten of the VBC-state in question are therefore given the same weight.

We use a VMC scheme, with the more efficient algorithm first introduced by Ceperly et al \cite{Cep1}, this algorithm computes the ratio of the desired determinants for each MC step, with the help of the inverse Matrix. We performed a few runs with the naive approach, computing directly both determinants to make sure the rounding error resulting from the improved algorithm is not causing any bias. We found that the usual updating scheme over n.n. neighbors converges slowly, particularly at half-filling. The reason is that the weights of the wave-function fluctuate strongly, causing the system to get trapped in certain configurations for a long time. Sampling over n.n. and n.n.n. neighbors improved the convergence. We compared the result of both versions for special cases and did not find any difference. For fractional fillings, it is hard to avoid degenerate states. As the result shows very little dependence on the boundary conditions, we chose to work with periodic boundary conditions (PBC). This choice is not ideal at e.g. half-filling, however comparing different geometries and fillings, we found it the most ``consistent'' choice. One sweep is defined as an update over all n.n. and n.n.n. neighbors. Typically we performed around 20.000 such sweeps to equilibriate the system and another 20.000 for measurements. We then average our data over at least 8 independent runs, and obtained an error-bar. To investigate the finite-size behavior we used lattices from $N=192$ to $N=432$ sites. Our larger sizes used, compare to the sizes currently the larger sizes for VMC on the Heisenberg-model \cite{RAN1,IQBAL}. The $t-J$ model is numerically about one magnitute harder. Because we used two physical geometries we will describe the system-size by $N$ not by $L$ to avoid confusion as $N=L^2 \times 3$ or $N=L^2 \times 9$ depending on the unit-cell in question. Our program is flexible in terms of wave-function and lattices, this gives us the possibility to check our code against published results.
To estimate the quality of our result we compare our result with the $N=21$ site ED-result of L\"auchli \cite{ANDED}. To optimize our variational parameter, we used the naive ``mesh'' approach. We are aware of more sophisticated methods such as the SR-method \cite{SOR}. The later method is undoubtedly very good for the optimization of wave-functions with several variational parameters, for this first exploratory investigation, we chose to keep the number of variational parameter small, for this reason the mesh-method is sufficient.

\section{Results}

\begin{figure}
     \centering
    \includegraphics[width=\columnwidth]{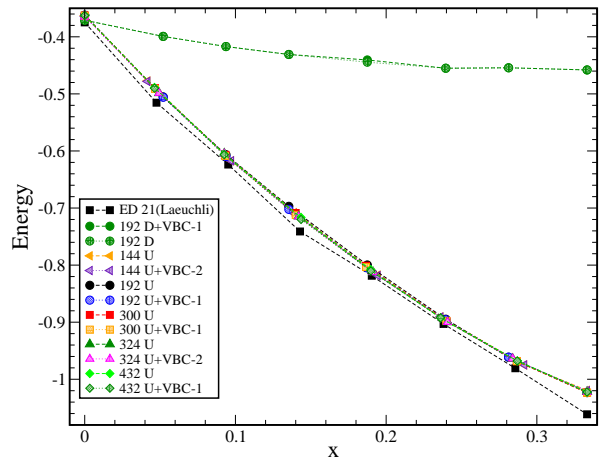}
    \caption{(Color online) Ground state energy for various states and system-sizes in their optimum as a function of doping $x$ and the comparison with the ED \cite{ANDED} on small clusters of $N=21$.
The error-bar is smaller than the symbol-size.}
    \label{result}
\end{figure}

\begin{figure}
    \centering
    \includegraphics[width=\columnwidth]{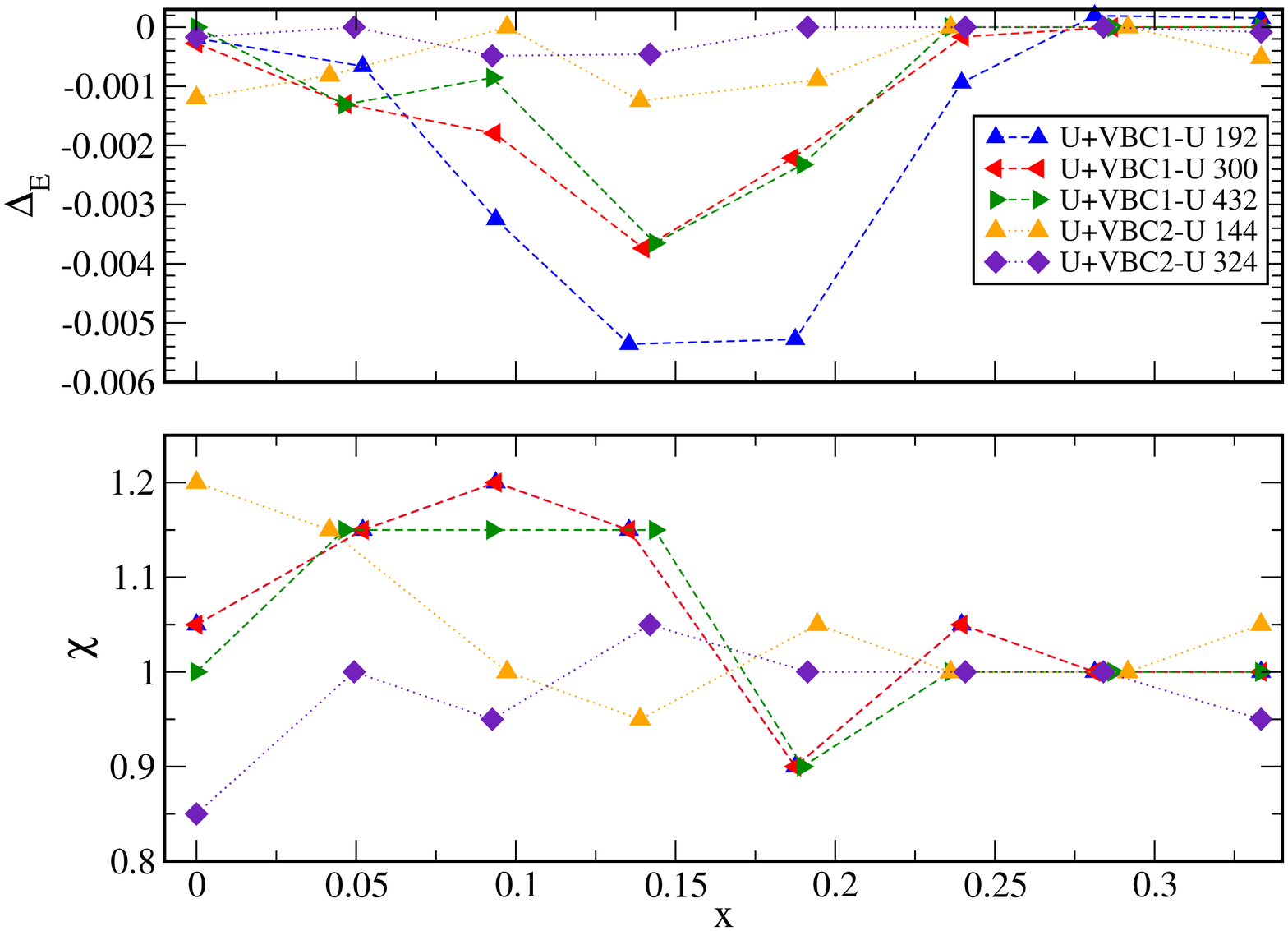}
    \caption{(Color online) Above: Difference between the U state energy compared to the U+VBC-1 state and U+VBC-2 state energies for various sizes. Below: Variational parameter $\chi$ (as described in the text) for the U+VBC-1 and U+VBC-2 states for some sizes as a function of doping $x$. For clearity we show only a few typical error-bars.}
    \label{DELTAE}
\end{figure}

In Fig. \ref{result} we show the optimized energies of the
most important trial states as a function of doping. At half filling we
recover the result of Ran et al \cite{RAN1} the D-state. Already at very
low doping (the lowest doping-level possible for a spin-balanced state and for the sizes investigated) the flux of this state vanishes, as can be seen by comparing the
variational energies of the U- and D-states. On top of the U-state the variational parameters for
the VBC-1 and VBC-2 patten increase and further lower the energy. The energy
difference between the pure U-state and the U+VBC-1 and U+VBC-2 states and the
variational parameter corresponding to the states in questions are plotted in
Fig. \ref{DELTAE}. Apparently the VBC-2 patten shows only a very small
energy-gain compared to the pure U-state and a relative big finite-size dependence. The energy difference essentially vanishes for 
$N=324$. The VBC-1 patten seems
to have a stable finite-size behavior starting from $N=300$. For higher
doping in all cases the VBC-patten vanishes. As there is no systematic study on
the ground state energy as a function of the doping published, we included some results
on small clusters obtained by exact diagonalization by L\"auchli et al
(see Fig. \ref{result}). Our result is remarkable close to the ED result for
$N=21$. The results from ED do not allow an extrapolation, it is therefore not
clear how well our energy agrees, yet the close match of the two methods which
is almost within 2 percent is very encouraging. At a filling of 2/3 our result
shows larger derivation from the ED result, suggesting that the real ground state is not described well with our wave-function starting from this doping level.\\

\begin{figure}
     \centering
    \includegraphics[width=\columnwidth]{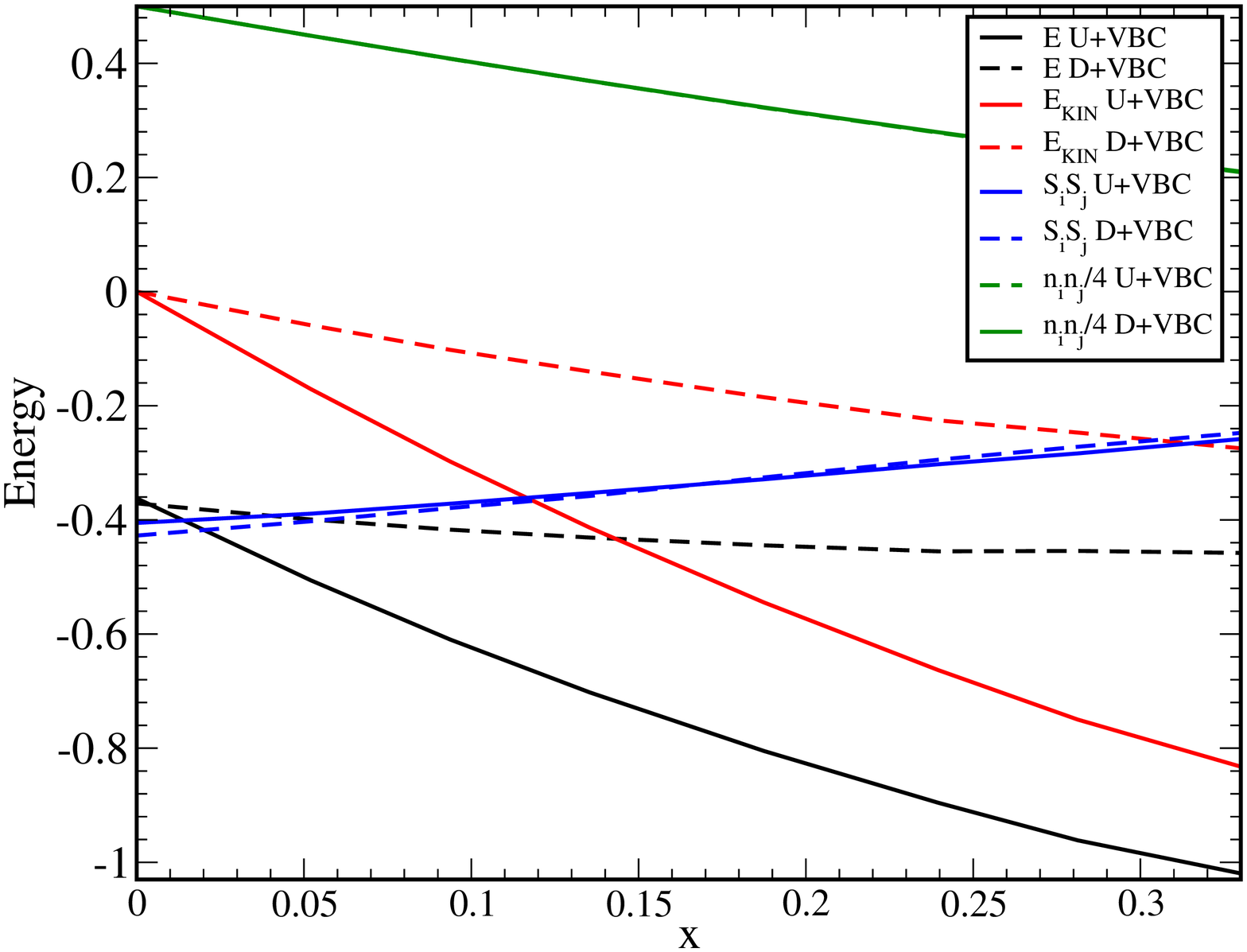}
    \caption{(Color online) Contributions of the various terms in the $t-J$
      model (kinetic energy, spin-exchange and $\frac{n_i n_j}{4}$-term, compare Eq. \ref{ham}), for the D+VBC-1 state and the U+VBC-1 state with the best energy at each doping-level $x$.}
    \label{CONTR}
\end{figure}

\begin{figure}
     \centering
    \includegraphics[width=\columnwidth]{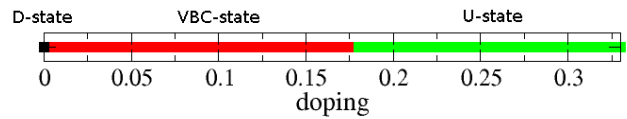}
    \caption{(Color online) Phase diagram of the $t-J$ model on the Kagome lattice.}
    \label{PHASE}
\end{figure}

In Fig. \ref{CONTR} the contributions of the 3 terms in the $t-J$ model for the
D+VBC-1 and U+VBC-1 states are plotted. As can be seen the kinetic energy
grows with the doping for the U-state and D-state, as we introduce holes. The
spin-exchange reduces as expected. This reduction is relatively small compared
for the introduced doping. At half filling the D-state has a slightly lower
spin-spin exchange. At a doping level above $x=0.18$ the U-state spin-spin exchange
energy starts to be lower compared to the one of the D-state.
The biggest difference between the two states comes
from the kinetic energy, which decreases much faster for the U-state compared
to the D-state. This is the reason why only at or extremely close to half-filling the D-state is the best variational state.
 The holes remain isolated as can be seen from the $n_i n_j$
term, indicating that no phase-separation is taking place. Apparently the
holes move in the background of an almost ideal VBC patten for the doping-levels it is
formed. The low value of the bond-order weight as a variational parameter
seems to be necessary to keep the holes mobile. From the data we can infer the two
competing trends of hole-mobility and dimerization. Thus in agreement with
earlier ED studies of static holes and single mobile holes, and with the
results obtained for 2/3 filling of the $t-J$ model. Introducing chirality (C-state) as described above increases the energy. The U+VBC-2
state gives almost the same energy compared to the pure U-state. The D+VBC-2 state gives very similar energies
as the D+VBC-1 state (which by itself does not differ much from the pure
D-state). These states are not shown here. The variational parameter for the bond-amplitude develops a maximum
for a  doping of around $x=0.09$ then the parameter decreases again. The maximum for the
energy-difference between the U-state and the U+VBC-1 state is at around
$x=0.14$. Considering the results for the energies and the variational
parameter, we suggest a phasediagram as depicted in Fig. \ref{PHASE}.

To summarize, we have for the first time studied the $t-J$ model on the Kagome lattice 
systematically for low doping. Our results are consistent with present ED
on the $N=21$ site cluster. We find that the spin-liquid state found for half-filling is
unstable even for very small doping and a zero-flux state is formed. This state shows
VBC tendencies which vanish for larger doping. There are a number of interesting question remaining for future studies: most promising are a more free parametrisation for the VBCs and the introduction of pairing channels.\\

{\it Acknowledgements: } We wish to thank Andreas L\"auchli for providing ED results for the $N=21$ cluster. Initially ALPS 2.0 has been used to perform exact diagonalization on small clusters (12- and 18-sites) to test our results \cite{ALPS}. 
The VMC calculations have been carried out on the
HPC facilities of the RWTH-Aachen. SG wants to thank for the technical support at this facility. Discussions with 
Andreas L\"auchli and Jung Hong Han are acknowledged.

\end{document}